\documentclass[12pt]{article}
\pdfoutput=1
\usepackage{latexsym}
\usepackage{epsfig,amssymb,euscript}
\usepackage{amsmath} 
\usepackage{slashed}
\usepackage{color}
\usepackage{cite}
\usepackage{graphicx}
\usepackage{hyperref}
\usepackage{subdepth}
\usepackage[margin=3cm]{geometry}

\numberwithin{equation}{section}

\newcommand{\nn} {\nonumber\\}

\def\tilde{\widetilde}

\def\sfU{\mathsf{U}}

\def\ImO{\mathrm{Im}\mathcal{O}}
\def\ImOt{\mathrm{Im}\tilde{\mathcal{O}}}

\catcode`@=12

\def\g{\gamma}

\def\n{\nu}
  
\def\pt{\tilde{\varphi}}

\def\6{\partial}

\def\rt{\tilde{\rho}_0}
\def\pt{\tilde{\pi}_0}

\begin{document}
\begin{titlepage}

\begin{center}
\parbox[t]{\textwidth}{\centering \LARGE	\bf
	\vspace{1em}
	Holographic Ward identities\\
	{for}
	symmetry breaking in two dimensions
	}

\vspace{2em}

{\large 
		Riccardo Argurio,$^{1}$ Gaston Giribet,$^{2,3}$\\ 
		\vspace{1ex}Andrea Marzolla,$^{1}$ Daniel Naegels$^{1}$ and J. Anibal Sierra-Garcia$^{4}$}

\vspace{2em}
\parbox[t]{0.85\textwidth}{\itshape \linespread{1} \footnotesize \centering
{$^1$ Physique Th\'eorique et Math\'ematique and International Solvay Institutes,\\ Universit\'e Libre de Bruxelles, C.P. 231, 1050 Brussels, Belgium.}

\vspace{1ex}
{$^2$ Martin Fisher School of Physics, Brandeis University, Waltham, Massachusetts 02453, USA.}

\vspace{1ex}
{$^3$ Physics Department, University of Buenos Aires FCEN-UBA and IFIBA-CONICET, Ciudad Universitaria, Pabell\'{o}n I, 1428, Buenos Aires, Argentina.}

\vspace{1ex}
{$^4$ Department of Particle Physics and IGFAE, University of Santiago de Compostela
E-15782 Santiago de Compostela, Spain.}
}

\end{center}

\vspace{3em}

\begin{center}
\bf \small Abstract
\end{center}

We investigate symmetry breaking in two-dimensional field theories which have a holographic gravity dual.  
Being at large $N$, the Coleman theorem does not hold and Goldstone bosons are expected. We consider the minimal setup to describe a conserved current and a charged operator, and we perform holographic renormalization in order to find the correct Ward identities describing symmetry breaking. This involves some subtleties related to the different boundary conditions that a vector can have in the three-dimensional bulk. We establish which is the correct prescription that yields, after renormalization, the same Ward identities as in higher dimensions.

\end{titlepage}

\tableofcontents

\section{Introduction}

The goal of this note is to investigate the issue of conserved currents and symmetry breaking in two-dimensional field theories which have a holographic gravity dual. This typically involves effects of the large $N$ limit on the field theory side.

In holography \cite{Maldacena:1997re,Witten:1998qj,Gubser:1998bc}, symmetries and their breaking have been discussed early on \cite{Klebanov:1999tb,Bianchi:2001de,Bianchi:2001kw}. Most of the discussion however has been done in boundary dimension higher than two. Indeed, on the field theory side, the established theorems on symmetry breaking~\cite{Goldstone:1962es} fail to apply to two-dimensional relativistic field theories~\cite{Coleman:1973ci}. This is usually referred to as the Coleman theorem, whose main point is that symmetry breaking necessarily implies a massless scalar boson, which in two-dimensions is ``ill-defined'' because of its non-decaying propagator in position space. A more physical intuition of this phenomenon goes as follows: 
{the long-range fluctuations of the gapless mode affect the order parameter, eventually setting the vacuum expectation value to zero~\cite{Ma:1974tp}.}
A similar story happens in one dimension higher at finite temperature~\cite{Mermin:1966fe,Hohenberg:1967zz}.

Some two-dimensional field theories can however behave in a way which is much reminiscent of symmetry breaking. This is particularly true for families of theories that have a large $N$ limit \cite{Coleman:1974jh,Gross:1974jv,Witten:1978qu}. In those theories, the decay of the order parameter correlations is suppressed by $1/N$, and in the strict infinite limit the order is restored. It is in this sense that one does not expect to see anything special in holography concerning symmetry breaking for boundary dimension two with respect to higher dimensions. A sign of the Coleman-Mermin-Wagner-Hohenberg theorem should only appear when $1/N$ corrections are taken into account. This is indeed what was found in \cite{Anninos:2010sq}, considering bulk tadpole diagrams for a three-dimensional boundary theory at finite temperature.

In this note we will consider a fixed AdS$_3$ background with a vector field and a complex scalar. This is the minimal bulk field content to describe a symmetry current and an operator charged under it. We will perform holographic renormalization in order to find the correct Ward identities describing symmetry breaking. This has a subtlety related to the fact that a vector in a three-dimensional bulk can have different boundary conditions~\cite{Marolf:2006nd}. We will see that coupling it to a charged scalar with a non-trivial profile singles out one boundary condition, the one that correctly corresponds to a conserved current in the boundary theory. Besides reproducing the Ward identities, including the situation where also explicit breaking is present, we will also analytically find the Goldstone boson in a specific toy example. 
{We will finally conclude with comments on quantum corrections, both in holography and in field theory.}


\section{Preamble: a free vector alone in AdS$\boldsymbol{_{3}}$}
The procedure of holographic renormalization exhibits many subtleties {for a vector} in 2+1 bulk dimensions. Most of them are related to the fact that a vector in AdS$_3$ has analogous properties as a scalar at the Breitenlohner-Freedman (BF) bound \cite{Breitenlohner:1982jf}. We then start the study of the peculiarities of holographic renormalization in two dimensions with a preliminary discussion of a free gauge field in AdS$_3$, before coupling it to matter and analyzing the physics of symmetry breaking.

As mentioned in the introduction, the case of a vector in AdS$_3$ was discussed briefly in \cite{Marolf:2006nd} along with higher dimensions. It was stated that only a particular boundary condition led to normalizable fluctuations. It was also noted there that it can be useful to dualize the bulk vector into a bulk massless scalar. This approach was also followed in \cite{Faulkner:2012gt}, where different boundary conditions were selected, in the dual frame. 

{Here, we will stick to a vector in bulk AdS$_3$ and discuss how one has to perform renormalization when imposing 
the boundary condition of \cite{Marolf:2006nd}. In the next sections we will show that a different boundary condition is needed for consistency with the derivation of the Ward identities.}

We take the following bulk action for a free Abelian gauge field:\footnote{{Being in three dimensions, one could include a Chern-Simons term for the vector (see for instance \cite{Andrade:2011sx} for a careful discussion in a similar perspective). Since our aim is to stay as close as possible to the higher dimensional cases, where such a term is not present, we will take here the minimalistic approach and set it to zero. This choice is of course protected by parity.}}
\begin{equation}	\label{Svec}
S=\int\!d^3x \sqrt{-g} \left( -\dfrac{1}{4}F^{MN}F_{MN} \right) \ ,
\end{equation}
where $F_{MN}$ is the usual electromagnetic field strength, and $g_{MN}$ is the AdS$_3$ metric in the Poincar\'e patch,
\begin{equation*}
g_{MN}dx^Mdx^N = \frac{1}{z^2}\left(dz^2-dt^2+dx^2\right) \ .
\end{equation*}
We choose the radial gauge~$A_z\!=\!0$, and we divide the remainder into transversal and longitudinal components,
\begin{equation}	\label{AT+iL}
A_{\mu}=T_{\mu}+\partial_{\mu}L \ , \quad \text{with }\ \partial_{\mu}T^{\mu}=0 \ ,
\end{equation}
so that the action becomes
\begin{equation}	\label{Strv}
S=-\int\!d^3x\; \frac{z}{2}\Big[-\partial_z L\Box\partial_z L +\partial_zT^{\mu}\partial_zT_{\mu} - T^\mu\Box T_\mu \Big] \ .
\end{equation}
{The action \eqref{Svec} leads to the following equations of motion:}
\begin{align}
&	\Box\partial_zL =0 \ , \phantom{\big]} \label{eqvecAz}\\
&	z\partial_z(z\partial_zL) =0  \ , \phantom{\big]} \label{eqvecL}\\
& 	z\partial_z(z\partial_zT_{\mu}) +z^2\Box T_{\mu} =0 \ . \phantom{\big]} \label{eqvecT}
\end{align}
From the last two equations we derive the asymptotic behaviors of the fields,
\begin{equation}	\label{asymvec}
\begin{aligned}
L = \ln\!z\,\tilde{l}_{0} +l_{0} +\:\ldots \ ,	\qquad
T^{\mu} = \ln\!z\,\tilde{t}^\mu_{0} +t^\mu_{0} +\:\ldots \ ,	
\end{aligned}
\end{equation}
while from the first one we can drop the term in the action \eqref{Strv} involving the longitudinal component
. {Note that the presence of the logarithmic terms entails that the constant terms can suffer from an ambiguity. We will deal later on with this issue.}

Then we can integrate by parts and use the equation of motion for $T_\mu$ to express the action as a boundary term:
\begin{equation}\label{la28}
S_{reg}=\frac{1}{2}\int_{z=\epsilon} d^{2}x\; T^{\mu}z\partial_zT_{\mu} = 
	\frac{1}{2}\int_{z=\epsilon} d^{2}x\; \big(\ln\!z\,\tilde{t}_0 +t_0\big)\cdot\tilde{t}_0 \ ,
\end{equation}
which needs to be renormalized because of the logarithmic divergence.

If we want to write a counterterm that removes this divergence and is gauge invariant, we may build it out of the field strength, but we soon realize that we then have to make it non local. This turns out to be equivalent to a mass term, which is absolutely local, but gauge invariant only on-shell, by equation~\eqref{eqvecAz}. Indeed,\footnote{{A counterterm with a $1/\ln\!z$ prefactor is typically needed for scalars at the BF bound, see e.g.~\cite{Bianchi:2001kw}.}}
\begin{align}	\label{Svec_ct}
S_{ct} &= 
	-\dfrac{1}{4}\int\!d^2x\; \frac{\sqrt{-\gamma\,}}{\ln\!z\,\gamma^{\rho\sigma}\partial_\rho\partial_\sigma}\, \gamma^{\kappa\mu}\gamma^{\lambda\nu} F_{\kappa\lambda}F_{\mu\nu} = \nn 
&	=
	\dfrac{1}{2}\int\!d^2x\;\frac{1}{\ln\!z}\, T_\mu T^\mu = \dfrac{1}{2}\int\!d^2x\;\frac{\sqrt{-\gamma\,}}{\ln\!z}\ \gamma^{\mu\nu}A_\mu A_\nu = \nn
&	=
		\frac{1}{2}\int_{z=\epsilon}\!d^{2}x\; \big(\ln\!z\,\tilde{t}_0 +2t_0\big)\cdot\tilde{t}_0 \ , 
\end{align}
where $\gamma_{\mu\nu}$ is the induced metric on the two-dimensional boundary, and the identity in the second line holds indeed thanks to the constraint~\eqref{eqvecAz}.

With such counterterm the renormalized action, $S_{ren}\!=\!S_{reg}\!-\!S_{ct}$, reads
\begin{equation}	\label{SrenT}
S_{ren} = -\frac{1}{2}\int_{z=\epsilon} \!d^{2}x\;\; \tilde{t}_0 \cdot t_0 \ .
\end{equation}

In holography, the source of the dual operator is defined as the mode that has to be fixed in order to satisfy the variational principle. 
We then take the variation of the bulk action~\eqref{Svec} with respect to the fields and we put it on-shell, obtaining
\begin{equation}
\delta S_{on-shell} = \int_{z=\epsilon}\!d^{2}x\; \Big[\;
	\delta T^{\mu}z\partial_{z}T_{\mu} -\delta L\Box z\partial_zL \Big] 
=\int_{z=\epsilon}\!d^{2}x\; 
	\tilde{t}_{0}\cdot\big(\ln\!z\,\delta\tilde{t}_{0}+\delta t_{0}\big)\ .
\end{equation}
We variate the counterterm~\eqref{Svec_ct} as well, and we eventually get the variation of the renormalized action
\begin{equation}	\label{varttil}
\delta S_{ren}=\delta S_{on-shell}-\delta S_{ct} = -\int_{z=\epsilon}\!d^{2}x\; 
	{t}_{0}\cdot \delta\tilde t_{0}\ .
\end{equation}
Hence we see that we have to fix $\tilde{t}_0^\mu$ in order to satisfy the variational principle, and so the source for the operator dual to $A^\mu$ is the coefficient of the logarithm, in agreement with~\cite{Marolf:2006nd}. Note that as the source is transverse, the dual operator enjoys a gauge symmetry and thus cannot be a conserved current: it would be a pure gauge field. 

We stress that the counterterm we have to introduce in two boundary dimensions, which has the form of a mass term, does not have an equivalent in any higher dimensions. This might be reminiscent of the Schwinger model (see e.g.~\cite{Gross:1995bp} for a modern exposition),
{where the photon mass is also generated by exactly the same non-local term.} Note however that here we are dealing with a non-local counterterm, due to a non-local UV divergent term, while in two-dimensional QED the loop-generated mass of the photon is finite. 

In \cite{Marolf:2006nd} it is noted that the above boundary conditions for the vector are equivalent to the usual boundary conditions one would impose on the massless scalar that is equivalent to (the transverse part of) the vector by bulk duality:
\begin{equation}
\partial_L\vartheta\!=\!\sqrt{-g\,}g^{MR}g^{NS}\varepsilon_{LMN}\partial_R A_S\ .
\end{equation}
It is straightforward to see that the usual, local counterterm that one writes for $\vartheta$ corresponds to the non-local one found above. 

Profiting from this dual formulation, the authors of \cite{Faulkner:2012gt} have argued that in order to describe a conserved current in the boundary theory, one should impose mixed boundary conditions on $\vartheta$, which are interpolating between the ordinary and the alternative quantizations \cite{Klebanov:1999tb} (see also \cite{Minces:1999eg}).\footnote{
	We should note that when considering the holographic correspondence between string theory on AdS$_3$ and boundary CFT$_2$ as in \cite{Giveon:1998ns}, there is a natural prescription to describe CFT currents, {which are actually enhanced to Kac-Moody generators.} The techniques are however different from the ones employed here, in particular there is no renormalization.}
Below, we are going to see that coupling the vector to a scalar leaves us with the only choice of the alternative quantization for the vector. 

\section{Holographic renormalization {with} a charged scalar}
We consider now a holographic model for spontaneous symmetry breaking in a 1+1~dimensional boundary field theory. We thus consider the following action, of an Abelian gauge field coupled to a complex scalar in AdS$_3$:
\begin{equation}
S=\int\!d^3x \sqrt{-g} \left[ -\dfrac{1}{4}F^{MN}F_{MN}-g^{MN}(D_{M}\phi)^{*}D_{N}\phi-m_{\phi}^{2}\,\phi^{*}\phi \right] \ , 
\label{S-U1}
\end{equation}
where
\begin{equation*}
\begin{aligned}
&	F_{MN} =	\partial_{M}A_{N}-\partial_{N}A_{M} \ ,\\
&	D_{M} =	\partial_{M}-iA_{M} \ .
\end{aligned}
\end{equation*}
From the equations of motion of a free scalar in AdS$_3$, one can infer the following scaling dimensions for the dual boundary operator/source:
\begin{equation}
\Delta_\pm = 1\pm\nu\ , \quad \text{with }\ \nu = \sqrt{1+m_{\phi}^2}\ .
\label{Delta+-}
\end{equation}
We then fluctuate the complex scalar around a fixed background,
\begin{equation}
\phi= \frac{\phi_B+\rho+i\pi}{\sqrt{2}}\ , \quad \text{with }\ \phi_B=m\,z^{1-\nu}+v\,z^{1+\nu} \ .
\end{equation}
We take $m$ and $v$ to be real for definiteness. When the scalar is in the ordinary quantization, the sub-leading piece (proportional to~$v$)  triggers a VEV for the real part of the dual boundary operator, and so  leads to spontaneous symmetry breaking of the global $\mathrm{U}(1)$. The leading piece (proportional to~$m$)  corresponds to explicit breaking of the symmetry, and we keep it different from zero for the moment, in order to use it as a sort of regulator. It will indeed turn out that we will  need it in some intermediary steps. 

Moreover, we fix the radial gauge $A_z=0$ and we conveniently split the gauge field into transverse and longitudinal components as in~\eqref{AT+iL}. We then derive from the variation of the action the following linearized equations of motion for the fluctuated fields:
\begin{align}
&	\Box z^2\partial_zL -\left(\phi_B\partial_z\pi-\pi\partial_z\phi_B\right) =0 \ , \phantom{\frac{}{|}} \label{eqAz}\\
&	z^2\partial_z^2T_{\mu} +z\partial_zT_{\mu} +z^2\Box T_{\mu} -\phi_B^2T_{\mu} =0 \ , \phantom{\frac{}{|}} \label{eqAt}\\
&	z^2\partial_z^2L +z\partial_zL -\phi_B^2 L +\phi_B\pi =0 \phantom{\frac{}{|}} \ , \label{eqL}\\
&	z^2\partial_{z}^{2}\rho -z\partial_{z}\rho -m^{2}\rho +z^2\Box\rho =0 \phantom{\frac{}{|}} \ , \label{eqrho}\\
&	z^2\partial_{z}^{2}\pi -z\partial_{z}\pi -m^{2}\pi +z^2\Box\pi -z^2\phi_B\Box L =0 \ . \label{eqpi}
\end{align}

As we have seen in the previous section, in three dimensions the vector field is at the BF bound, and indeed we have the following asymptotic expansions near the boundary:
\begin{equation}
T^{\mu}	= \ln\!z\,\tilde{t}_0^\mu +t_0^\mu +\:\ldots \ , \qquad
	L = \ln\!z\,\tilde{l}_{0} +l_0 +\:\ldots \ .		\label{asymL}
\end{equation}
The asymptotic expansion of the two scalar components depends on the value of the bulk mass. Let us set ourselves in the window between the BF bound~($m^2\!=\!-1$) and the ``massless bound''~($m^2\!=\!0$), and exclude the two extremal values, which would need to be treated separately since they entail logarithms. For all other values in this window, the scalar asymptotic expansions are logarithm-free. We thus have the following expansions,
\begin{equation}
\begin{aligned}
\rho &=	z^{1-\nu}\left(\rho_0+z^2\rho_1+\,\ldots\right) +z^{1+\nu}\left(\tilde{\rho}_0 +z^2\tilde{\rho}_1 +\,\ldots\right) \ , \phantom{\frac{}{|}}\\
\pi  &=	z^{1-\nu}\left(\pi_0+z^2\pi_1+\,\ldots\right) +z^{1+\nu}\left(\tilde{\pi}_0 +z^2\tilde{\pi}_1 +\,\ldots\right)\ ;  \phantom{\frac{|}{}}
\end{aligned}\ \quad \text{for }\ \nu \in\; ]0,1[ \ .
\end{equation}

We can now, integrating by parts and using the equations of motion, reduce the action to a boundary term, which reads
\begin{align}
S_{reg} & = \int_{z=\epsilon}\!\!d^{2}x\; \frac{1}{2} \bigg[\,
	T^{\mu}z\partial_zT_{\mu} -\Box Lz\partial_zL +\frac{1}{z}\left(\pi\partial_z\pi+\rho\partial_z\rho+2\rho\partial_z\phi_B\right) \bigg]\ ,
\label{SregGenCase}
\end{align}
where we neglect the terms at the zeroth order in the fluctuations, which are not relevant to our discussion. By using the asymptotic expansions we obtain
\begin{align}	\label{regActionAsymptoticGenCase}
S_{reg} = \int_{z=\epsilon}\!\! d^{2}x\; \frac{1}{2} \Big[ &\;
	\big(\ln\!z\,\tilde{t}_0 +t_0\big)\cdot \tilde{t}_0 -\big(\ln\!z\,\tilde{l}_0 +l_0\big)\,\Box\tilde{l}_0\; + \nn
&
	+\rho_0\,\Big( (1-\nu)\big(\rho_0+2m\big)z^{-2\nu} +2\tilde{\rho}_0 \Big) +2m(1-\nu)\,\tilde{\rho}_0 +2v(1+\nu)\,\rho_0 + \nn
&
	+\pi_0\,\Big((1-\nu)\pi_0\,z^{-2\nu} +2\tilde{\pi}_0\Big) \Big] \ . 
\end{align}

We see that the divergent pieces of the scalar sector can be removed by the usual counterterm (in which we subtract the background value)
\begin{align}\label{Sctmass}
S^{(m)}_{ct} &= 
	\left(1-\nu\right)\int_{z=\epsilon}\! d^{2}x\,\sqrt{-\gamma\,}\;\left(\phi^*\phi-\frac{\phi_B^2}{2}\right) \\
&	= \frac{1}{2}\left(1-\nu\right)\int_{z=\epsilon}\! d^{2}x\,\sqrt{-\gamma\,}\; \Big[\rho^2 +2\phi_B\rho +\pi^2 \,\Big] \ , 	\nonumber
\end{align}
leaving only the logarithmic divergences of the vector sector:
\begin{align}	\label{Sreg-Sctm}
S_{reg}-S^{(m)}_{ct} = \frac{1}{2}\int\!d^{2}x\; \Big[&	
	\big(\ln\!z\,\tilde{t}_0 +t_0\big)\cdot \tilde{t}_0 -\big(\ln\!z\,\tilde{l}_0 +l_0\big)\,\Box\tilde{l}_0 \:+ \nn
&\qquad\qquad\qquad
	 +2\nu\, \big(\rho_0\tilde{\rho}_0 +2v\,\rho_0 +\pi_0\tilde{\pi}_0\big)\Big] \ . 
\end{align}


We would like to remove also these divergences and then express the renormalized action in terms of the sources only. To do so, we need to identify which are the sources. Normally the source is associated with the leading term in the small~$z$ expansion of the field, unless one performs alternative quantization. For the gauge vector field, in any higher dimension the leading piece corresponds to a constant term. In our case instead, due to the fact that the vector is at the BF bound and to the consequent presence of the logarithmic term, the leading term is no longer the constant one. Then we would be naively led to choose the logarithmic term as the source, as we indeed did in the previous section. 
For the transverse part this does not pose any particular problem, but for the longitudinal part, which in the present case does not disappear from the boundary action, it is more problematic. 

Note that the longitudinal part of the vector shifts under gauge transformations, which in the radial gauge~$A_z\!=\!0$ are constant in $z$. It is then the constant term in $L$ that shifts, in any boundary dimensions, including two. Thus in our case it is $l_0$ that shifts under gauge transformations. In other words, it is the constant part of $A_\mu$ that has gauge transformations, and so should be the source of a boundary conserved current.
This seems then reasonable to try to alternatively quantize the vector, and move the source from the coefficient of the logarithm to the $z$-constant, gauge-dependent term.

\subsection{Ordinary quantization for the vector}

In first place, however, let us renormalize the vector in ordinary quantization,  {i.e.~keeping the source to be $\tilde l_0$. The point of this subsection is to show that this approach does however lead to a flawed physical picture, and that a different choice has to be made.}

We then compute first the variation of the action~\eqref{S-U1}, putting on-shell the bulk part as usual:
\begin{align}\label{deltaSreg}
\delta S_{on-shell}=\int_{z=\epsilon}\!d^{2}x\; \Big[\;&
	\delta T^{\mu}z\partial_{z}T_{\mu} -\delta L\Box z\partial_zL +\frac{1}{z}\left(\delta \pi\partial_z \pi+\delta\rho\partial_z\rho+\delta\rho\partial_z\phi_B\right) \Big] \nn
=\int_{z=\epsilon}\!d^{2}x\; \Big[\;& 
	\tilde{t}_{0}\cdot\big(\ln\!z\,\delta\tilde{t}_{0}+\delta t_{0}\big) -\Box\tilde{l}_0\big(\ln\!z\,\delta\tilde{l}_0 +\delta l_0\big) \,+ \nn
\phantom{\Big|}& 
	+(1-\nu)\pi_{0}\big(z^{-2\nu}\delta\pi_0+\delta\tilde{\pi}_0\big) +(1+\nu)\tilde{\pi}_0\delta\pi_0 \,+ \nn
& 
	+(1-\nu)\big(\rho_{0}+m\big)\big(z^{-2\nu}\delta\rho_{0}+\delta\tilde{\rho}_{0}\big) +(1+\nu)\big(\tilde{\rho}_{0}+v\big)\delta\rho_{0} \,\Big]\ .
\end{align}
It is crucial here not to use the constraint~\eqref{eqAz}, which in components yields
\begin{equation}
\Box\tilde{l}_0 = 2\nu\left(m\tilde{\pi}_0-\,v\,\pi_0\right) \ . 		\label{boxtill}
\end{equation} 
This equation 
is relating the coefficient of the logarithm to the source and vev of the fluctuating scalar $\pi$. The equations of motion can be used to express vev's in term of sources, but since we do not know yet whether $\tilde{l}_0$ will be a source or not, we have to remain off-shell to check the variational principle.\footnote{Using \eqref{boxtill}  to remove $\tilde{l}_0$ from the action would actually imply that it is not a source, forcing this choice from the start.}

We then vary the counterterm for the scalar divergences,
\begin{align}
\delta S_{ct}^{(m)} = (1-\nu)\!\int_{z=\epsilon}\!d^{2}x\, \Big[\,& 
	\Big( z^{-2\nu}\big(\rho_{0}+m\big)+\big(\tilde{\rho}_{0}+v\big) \Big)\delta\rho_0 +\big(\rho_{0}+m\big)\delta\tilde{\rho}_{0} \nn
&	
	+\Big( z^{-2\nu}\pi_{0} +\tilde{\pi}_{0} \Big)\delta\pi_{0} +\pi_{0}\delta\tilde{\pi}_{0} \Big] \ ,
\end{align}
so that we get
\begin{align}\label{deltaSreg-Sctm}
\!\delta S_{on-shell}- \delta S_{ct}^{(m)} = \int_{z=\epsilon}\!d^{2}x\; \Big[\;&
	\tilde{t}_{0}\cdot\big(\ln \!z\delta\tilde{t}_{0}+\delta t_{0}\big) -\Box\tilde{l}_0\big(\ln\!z\,\delta\tilde{l}_0 +\delta l_0\big) \,+ \nn
&\;
	+2\nu\:\Big( \big(\tilde{\rho}_{0}+v\big)\delta\rho_0 +\tilde{\pi}_0\delta\pi_{0} \Big) \,\Big]\ .
\end{align}
Thus the scalar sources appear to be well-defined in the ordinary quantization, but of course we still have to renormalize the vector sector. A mass-like counterterm as the one of the previous section \eqref{Svec_ct} will not help in renormalizing the longitudinal component. We propose the following gauge invariant, local counterterm:
\begin{align}\label{Sct0}
S_{ct}^{(0)} & = \int_{z=\epsilon}\! d^{2}x\,\frac{\sqrt{-\gamma\,}\,\gamma^{\mu\nu}}{\ln\!z\;\phi_B^2}\: (D_\mu\phi)^*D_\nu\phi \\
&	=
	\frac{1}{2}\int_{z=\epsilon}\! d^{2}x\;\bigg[ \left(\ln\!z\,\tilde{t}_0 +2t_0\right)\cdot\tilde{t}_0 -\Big(\ln\!z\,\tilde{l}_0 +2l_0 -\frac{2}{m}\pi_0\Big)\Box\tilde{l}_0 \,\bigg]\ ,	\nonumber
\end{align}
whose variation is
\begin{align}\label{varSct0}
\delta S_{ct}^{(0)} = \int_{z=\epsilon}\!d^{2}x\;\bigg[\:&
\big(\ln\!z\,\tilde{t}_0 +t_0\big)\cdot\delta\tilde{t}_0 +\tilde{t}_0\cdot\delta{t_0} \\
&
-\Box\tilde{l_0}\,\big(\ln\!z\,\delta\tilde{l}_0 +\delta{l_0}\big) -\Box{l_0}\delta\tilde{l_0} +\frac{1}{m}\left(\Box\pi_0\delta\tilde{l_0}+\Box\tilde{l_0}\delta\pi_0\right)\Big) \,\bigg]\ . 	\nonumber
\end{align}
We notice that such counterterm (and its variation as well) is singular in the limit~$m\!\to\!0$, and so in the purely spontaneous case it would not do the job. Keeping instead $m\!\neq\!0$, we obtain
\begin{equation}
\delta\tilde{S}_{ren}=\int_{z=\epsilon}\!\!d^{2}x\: \bigg[ -t_0\cdot\delta\tilde{t}_0 -\frac{\Box}{m}\big(\pi_0-ml_0\big)\delta\tilde{l_0} +\big(2\nu\tilde{\pi}_0-\frac{\Box}{m}\tilde{l_0}\big)\delta{\pi}_0 \; +2\nu\left(\tilde{\rho}_0+v\right)\delta\rho_0 \bigg]\ .
\end{equation}
We see that in this way the variational principle is well defined (even if still singular in~$m\!\rightarrow\!0$), and in particular $\tilde{l}_0$ should be considered as the source.\footnote{Notice that the mass parameter $m$ acts here as a regulator in the case of the purely spontaneous case. The structure of counterterms looks different in the case one fixes $m=0$ ab initio. Keeping $m$ in the computation is convenient as it enables us to have control in the regularization of the general case. In the next subsection, this will yield a result that is regular in the $m\to 0$ limit, cf. (\ref{SrenAQ}).}
Then we had better interpret the constraint~\eqref{boxtill} as an expression for $\tilde{\pi}_0$ in terms of the sources. In this point of view the renormalized action is
\begin{equation}
\tilde{S}_{ren}=\frac{1}{2}\int_{z=\epsilon}\!d^{2}x\: \Big[ -\tilde{t}_0\cdot t_0 -\big(\pi_0-ml_0\big)\frac{\Box}{m}\tilde{l}_0 +2\nu\Big(\rho_0\tilde{\rho}_0+2v\rho_0+\frac{v}{m}\,{\pi_0}^2\Big) \Big]\ .
\end{equation}
Again we see that all the terms involving the source of the imaginary part of the dual scalar operator explode for $m\!=\!0$. No theory of spontaneous breaking can be extracted out of this, which is consistent with the fact that the vev of the longitudinal component of the vector is gauge-dependent in this quantization. Moreover, in the purely spontaneous case the constraint coming from eq.~\eqref{eqAz} becomes
\begin{equation}
\Box\tilde{l}_0=-2\nu\,v\pi_0 \ ,
\end{equation}
which strengthens the idea that $\tilde{l}_0$ cannot be the source of the conserved current. Indeed, $\pi_0$ is the source for the imaginary part of the scalar, and so $\tilde{l}_0$ cannot be another source,\footnote{A similar situation happens in cascading solutions, as for instance in \cite{Bertolini:2015hua}.} and therefore we confirm that we are forced to do the alternative quantization on the vector. 

Let us then show how to alternatively quantize the vector field and put the source back to the gauge-dependent, $z$-constant term~$l_0$, as in any higher dimensions.

\subsection{Alternative quantization for the vector}

We should then try to renormalize in a different way, such that we move the source to the constant term.
This is achieved by considering an additional counterterm, of the Legendre transform kind, such as the following one:
\begin{align}\label{Sct1}
{S}_{ct}^{(1)} &=
	\frac{i}{2\sqrt{2}}\int_{z=\epsilon}\! d^{2}x\, \frac{\sqrt{-\gamma\,}}{\phi_B}\;\gamma^{\mu\nu} z\partial_zA_\mu \Big(D_\nu\phi-(D_\nu\phi)^*\Big)\ = \nn
&=
	\frac12\int_{z=\epsilon}\! d^{2}x\, \Big[\: \big(\ln\!z\,\tilde{t}_0 +t_0\big)\cdot\tilde{t}_0 -\Big(\ln\!z\,\tilde{l}_0 -\frac{1}{m}\left(\pi_0-ml_0\right)\Big)\,\Box\tilde{l}_0 \,\Big]\ .
\end{align}
Indeed its variation is 
\begin{align}\label{varSct1}
\delta{S}_{ct}^{(1)} = \frac{1}{2}\int_{z=\epsilon}\!d^{2}x\;\Big[\:&
\big(\ln\!z\,2\tilde{t}_0 +t_0\big)\cdot\delta\tilde{t}_0 +\tilde{t}_0\cdot \delta{t_0} \\
&
-\big(\ln\!z\,2\tilde{l}_0 +l_0\big)\,\delta\Box\tilde{l_0} -\Box\tilde{l_0}\delta{l_0} +\frac{1}{m}\left(\pi_0\delta\Box\tilde{l_0}+\Box\tilde{l_0}\delta\pi_0\right)\Big]\ , 	\nonumber
\end{align}
and, taking the expression~\eqref{deltaSreg-Sctm}, and the variation of the ordinary counterterm~\eqref{Sct0}, the combination
\begin{align}\label{varSrenAQ}
&\delta S_{on-shell}-\delta S_{ct}^{(m)}+\delta S_{ct}^{(0)}-2\delta{S}_{ct}^{(1)} = \nn
&\qquad\qquad\qquad =
	\int_{z=\epsilon}\!d^{2}x\;\Big[\: \tilde{t}_0\cdot\delta{t_0} -\Box\tilde{l}_0\delta{l_0} +2\nu\:\Big( \big(\tilde{\rho}_{0}+v\big)\delta\rho_0 +\tilde{\pi}_0\delta\pi_{0} \Big) \,\Big]\ ,
\end{align}
yields the desired switch of the sources. Furthermore, the corresponding renormalized action reads
\begin{equation}\label{SrenAQ}
{S}_{ren}=\frac{1}{2}\int_{z=\epsilon} \!d^{2}x\; \Big[ t_0\cdot\tilde{t}_0 +2\nu\, \Big(\, \rho_0\tilde{\rho}_0 +2v\,\rho_0 +\pi_0\tilde{\pi}_0 -\big(m\tilde{\pi}_0-v\pi_0\big)l_0\Big)\Big] \ ,
\end{equation}
where we have used the constraint~\eqref{boxtill} to remove $\tilde{l}_0$, which is not a source anymore. Notice that both~\eqref{varSrenAQ} and~\eqref{SrenAQ} are finite in the $m\!\to\!0$ limit.

This renormalized action is completely identical to those of higher space-time dimensions and gives the suitable Ward identities for a pseudo-Goldstone boson (see for instance \cite{Argurio:2015wgr} for the three-dimensional case). However, both counterterms we have used~\mbox{(\ref{Sct0}, \ref{Sct1})} are singular for $m\!=\!0$ (even if the final result is not), so our action cannot be renormalized in this way if we set $m\!=\!0$ from the beginning.

We realize however that there is another gauge invariant and local counterterm that yields the same action as $2{S}_{ct}^{(1)}-S_{ct}^{(0)}$ in one step, namely
\begin{align}\label{Sct2}
S^{(2)}_{ct}&=
\frac{1}{2}\int_{z=\epsilon}\!d^{2}x\,\sqrt{-\gamma\,}\,\gamma^{\mu\nu} \ln\!z\big(z\partial_zA_\mu\big)\big(z\partial_zA_\nu\big) \nn
&=
\frac{1}{2}\int_{z=\epsilon}\!d^{2}x \ln\!z\left({\tilde{t}_0}\cdot{\tilde{t}_0}-\tilde{l}_0\Box\tilde{l}_0\right) \ .
\end{align}
This counterterm, subtracted to~\eqref{Sreg-Sctm}, trivially gives the renormalized action~\eqref{SrenAQ}. The counterterm is then equivalent to the combination~$2{S}_{ct}^{(1)}-S_{ct}^{(0)}$, and is not singular when $m\!=\!0$. 

Note that this could be interpreted to mean that the limits $z\rightarrow0$ (high UV) and $m\rightarrow0$ (purely spontaneous breaking) do not commute at large $N$. In order to make explicit this non-commutativity of the two limits, we can write
\begin{equation*}
2S^{(1)}_{ct}-S^{(0)}_{ct} = S^{(2)}_{ct} +\frac{1}{2}\int_{z=\epsilon}\!d^{2}x\:\frac{1}{\ln\!z}\bigg[
	2\Box{l_0}\,\frac{\pi_0+\tilde{\pi}_0z^{2\nu}}{m+vz^{2\nu}} 
	-\Box\pi_0\,\frac{\pi_0+2\tilde{\pi}_0z^{2\nu}}{m^2+2mvz^{2\nu}+v^2z^{4\nu}} \bigg]\ .
\end{equation*}
We see immediately that if we take first the limit~$z\!\rightarrow\!0$ we get the desired counterterm~${S}_{ct}^{(2)}$, which is independent of~$m$ and so the subsequent $m\!\rightarrow\!0$ limit is ineffective; whereas if we take first the limit~$m\!\rightarrow\!0$, we have no singularities  thanks to $v\!\neq\!0$, but we have surviving divergences in~$z$ when we take the $z\!\rightarrow\!0$ limit afterwards, namely:
\begin{equation}
\Big[2S^{(1)}_{ct}-S^{(0)}_{ct}\Big]_{m=0} = S^{(2)}_{ct} +\frac{1}{2}\int_{z=\epsilon}\!d^{2}x\:\bigg[
\frac{2}{v}\Box{l_0}\,\pi_0z^{-2\nu} -\frac{1}{v^2}\Box\pi_0\,\big(\pi_0z^{-4\nu}+2\tilde{\pi}_0z^{-2\nu}\big) \bigg]\ .
\end{equation}
One could be tempted to interpret this as a signal of the impossibility of taking \mbox{$v\!\neq\!0$} and \mbox{$m\!=\!0$} at the same time, i.e.~no spontaneous symmetry breaking. However the counterterm \eqref{Sct2} is perfectly well-behaved on its own, and it can actually be written also for the vector alone, whereas this is not the case for the two intermediate steps~\mbox{(\ref{Sct0}, \ref{Sct1})}, which are actually even more singular when all the scalar background is taken to zero. We cannot thus exclude the counterterm~\eqref{Sct2}, and with it we have to allow for spontaneous symmetry breaking in two dimensions, confirming the expected holographic evasion of the Coleman theorem.

{The counterterm \eqref{Sct2} has an explicit $\ln\!z$ factor, leading to the possibility to add also a finite counterterm with a similar structure and an arbitrary prefactor. This is indeed what was analyzed in \cite{Faulkner:2012gt} in the dual frame, with the interpretation of the introduction of a double-trace current-current deformation and the consequence of a non-trivial RG flow. Here we note that such a finite counterterm would spoil the identification of $t^\mu_{0}$ and $l_0$ as sources, shifting them by an arbitrary amount linear in $t^\mu_{0}$ and $\tilde{l}_0$ respectively. In the following, we take the point of view that the ambiguity in the $\ln\!z$ has been fixed, and we have taken $t^\mu_{0}$ and $l_0$ to be our sources. It is indeed this prescription that allows us to find the expected Ward identities.}

One last comment about alternative quantization for the vector field is that, if one holds $t_0^\mu$ fixed and lets $\tilde{t}_0^\mu$ loose, then according to~\cite{Marolf:2006nd} the fluctuations are not normalizable. This seems the price to pay to describe in the boundary theory a proper conserved current, whose existence we have no reason to exclude for a two-dimensional CFT. In addition, let us say that while for scalars bulk non-normalizability is usually connected to boundary operators with dimension below the unitarity bound, in the present case we do not see what problematic scenario this non-normalizability would correspond to in the dual theory; on the contrary, we have shown that everything works as smoothly as in higher dimensions precisely when we choose the alternative quantization for the vector field. 

\section{Symmetry breaking in AdS$\boldsymbol{_{3}}$/CFT$\boldsymbol{_{2}}$}
Once we have obtained the renormalized action \eqref{SrenAQ}, then showing that the Ward identities are realized is systematic. It actually follows from the same arguments as in \cite{Argurio:2015wgr}. First we rewrite the action as
\begin{align}\label{SrenAQexp}
{S}_{ren}=\frac{1}{2}\int_{z=\epsilon}\!d^{2}x\; \Big[ \;&
	t_0\cdot\tilde{t}_0 + \\
&	\;
	+2\nu\, \Big(\, \rho_0\tilde{\rho}_0 +v\left(2\rho_0+2\pi_0l_0-ml_0l_0\right) +\big(\pi_0-m l_0\big)\big(\tilde{\pi}_0-vl_0\big) \Big)\Big] \ . \nonumber
\end{align}
Then we remark that the equations of motions and gauge invariance dictate the relations between vevs and sources to take the following form:
\begin{equation}
\tilde t_0^\mu = f_t(\Box)  t_0^\mu\ , 	\quad 
	\tilde \rho_0 = f_\rho (\Box) \rho_0 \ , \quad 
		\tilde{\pi}_0-vl_0 = f_\pi (\Box)\big(\pi_0-m l_0\big)\ ,
\end{equation}
where the $f$'s are typically non-local functions, obtained by solving the EOM with appropriate IR boundary  conditions (i.e.~in the deep bulk). 

Replacing in the action yields the generating functional for one- and two-point functions, depending explicitly on sources only:
\begin{align}\label{SrenAQfinal}
{S}_{ren}=\frac{1}{2}\int_{z=\epsilon}\!d^{2}x\; \Big[ &\; 
	t_0\cdot f_t(\Box) t_0 +2\nu\, \Big(\, v\left(2\rho_0+2\pi_0l_0-m\,l_0l_0\right) + \\ 
& 
	+\rho_0f_\rho (\Box) \rho_0 +\big(\pi_0-m l_0\big) f_\pi (\Box)\big(\pi_0-m l_0\big)\Big)\Big] \ . \nonumber
\end{align}
Given the usual dictionary, for instance
\begin{equation}
\langle\ImO\rangle = \frac{\delta {S}_{ren} }{\delta \pi_0}\ ,\qquad 
\langle \partial^\mu j_\mu \rangle =- \frac{\delta {S}_{ren} }{\delta l_0}\ ,
\end{equation}
we get for the correlators most relevant  to the Ward identities
\begin{equation}\label{WIv}
\begin{aligned}
\big\langle \ImO(x)\,\ImO(x') \big\rangle & = -i\,2\nu\: f_\pi (\Box)\;\delta(x-x')\ , \\
	\big\langle \partial_\mu j^\mu(x)\,\ImO(x') \big\rangle &= -i\,2\nu\: \big( m f_\pi(\Box) -v\big)\; \delta(x-x')\ .
\end{aligned}\\
\end{equation}
As in \cite{Argurio:2015wgr}, we can obtain directly the Goldstone boson pole in the purely spontaneous case from the above relations. In momentum space, relativistic invariance and the Ward identity force the mixed correlator to be
\begin{equation}
\big\langle j_\mu(k)\,\ImO(-k)\big\rangle = v\,\frac{k_\mu}{k^2}\ ,
\end{equation}
displaying the expected massless pole. Furthermore when turning on $m$, one can argue that $f_\pi$ has to have a pole with a mass square proportional to $m$. Hence also $f_\pi$ has a massless pole in the $m=0$ limit. We will not repeat these steps here because they are clearly independent of the dimension of space-time (see \cite{Argurio:2015wgr} for a detailed discussion). 
The Coleman theorem kicks in only after one considers (perturbative) quantum corrections due to the massless particle. Clearly holography does not capture such quantum corrections, which we then assume to be suppressed by the large $N$ limit. 

Hence, the main result of the present note has been obtained, namely to show what is the correct prescription for the boundary conditions and for the renormalization of the vector in order to obtain the expected Ward identities in the two-dimensional boundary field theory. This has been derived through a physically intuitive operational method, based on locality, gauge-invariance and the variational principle. Of course it can be discussed in a more formal way through the Hamiltonian formalism, analogously to what has been done recently for the AdS$_2$ case in \cite{Cvetic:2016eiv,Erdmenger:2016jjg}, with conclusions (counterterms built out of canonical momenta) in agreement with ours.

In the following section, in order to cover all possibilities (namely, all scalar operator dimensions between $0$ and $2$), we will briefly perform alternative quantization also in the scalar sector. 
Moreover, this will allow us to work out an analytic expression for $f_\pi$ for a specific value of the dimension of the dual boundary operator.


\section{Alternative quantization of the scalar}
Here, as we did for the vector field, the goal is to move the sources to the subleading terms for the scalar as well. That is, we are interested in considering $\rt, \pt$ as the sources. 

In order to change the boundary conditions we should consider a Legendre transformation of the scalar counterterm~\eqref{Sctmass}, that is
\begin{align}
\hat{S}^{(m)}_{ct}& = 
	\frac12\int_{z=\epsilon } d^2x\,\sqrt{-\g\,}\Big(\phi^* z \partial_z \phi + \phi^* z \partial_z \phi -\phi_Bz\partial_z\phi_B\Big) \\
& =
	\frac12\int_{z=\epsilon } d^2x\; \Big[(1-\nu)\big(\rho_0\left(\rho_0+2m\right) +\pi_0\pi_0\big)\, z^{-2\nu } +2\,\big(v\rho_0 + m\tilde{\rho}_0 +\rho_0\tilde{\rho}_0 +\pi_0\tilde{\pi}_0\big)\Big] \nonumber .
\end{align}
Then the following combination is free from scalar divergences,
\begin{align}
S_{reg}+S^{(m)}_{ct}-2\hat{S}^{(m)}_{ct} = \frac{1}{2}\int\!d^{2}x\; \Big[\; &
	\big(\ln\!z\,\tilde{t}_0 +t_0\big)\cdot\tilde{t}_0 -\big(\ln\!z\,\tilde{l}_0 +l_0\big)\,\Box\tilde{l}_0 + \nn
&\qquad\qquad
	 -2\nu\, \big(\rho_0\tilde{\rho}_0 +2m\,\tilde{\rho}_0 +\pi_0\tilde{\pi}_0\big)\Big] \ ,
\end{align}
and $v$ and $m$ have opposite meanings with respect to~\eqref{Sreg-Sctm}. We can verify by the variational principle that indeed the sources and vev's are switched. If we take the expression~\eqref{deltaSreg} and subtract the variation of the present countenterm, we obtain
\begin{align}\label{deltaSreg-Sctm2}
&	\delta{S}_{on-shell}+\delta{S}^{(m)}_{ct}-2\delta\hat{S}^{(m)}_{ct} = \vphantom{\frac{}{|}}\\
&\quad\; =
	\int_{z=\epsilon}\!\!d^{2}x\: \Big[\,\tilde{t}_{0}\cdot\big(\ln\!z\delta\tilde{t}_{0}+\delta t_{0}\big) -\big(\ln\!z\,\delta\tilde{l}_0 +\delta l_0\big)\Box\tilde{l}_0 -2\nu\,\Big(\big(\rho_0+m\big)\delta\tilde{\rho}_{0} +\pi_0\delta\tilde{\pi}_0 \Big)\Big]\ , \nonumber
\end{align}
as desired.

Then we use the counterterm~\eqref{Sct2} to remove the vector divergences as well, and we get the renormalized action where both the vector and the scalar are in the alternative quantization:
\begin{equation}
\hat{S}_{ren} = \frac{1}{2} \int_{z=\epsilon } d^2x\; \Big[\, t_0\cdot\tilde{t}_0 -2\n\,\Big( \rho_0\rt +2m\rt +\pi_0\pt +\left(m\pt-v\pi_0\right) l_0\Big) \Big]\ .
\end{equation}

We remark that, since now the purely spontaneous breaking occurs for $v\!=\!0$, the two counterterms~\mbox{(\ref{Sct0}, \ref{Sct1})} are now well behaved for the purely spontaneous case, whereas they are singular for the purely explicit one. Since we do not expect any obstruction for explicit symmetry breaking specific to two dimensions, this confirms that the counterterm~\eqref{Sct2} is the correct one, while it is the ordinary quantization for the vector which is problematic.

If we now express the vev's in terms of the gauge-invariant sources in the following way:
\begin{equation}
\tilde{t}_0^\mu = f_t(\Box) t_0^\mu\ , 	\quad 
\rho_0 = \tilde{f}_\rho(\Box) \tilde{\rho}_0 \ , \quad 
\pi_0-ml_0 = \tilde{f}_\pi(\Box)\big(\tilde{\pi}_0-vl_0\big)
\end{equation}
(where the $\tilde{f}$'s are just the reciprocals of the $f$'s), we can rewrite the renormalized action uniquely in terms of the sources,
\begin{align}\label{hatSrenfinal}
\hat{S}_{ren}=\frac{1}{2}\int_{z=\epsilon}\!d^{2}x\; \Big[ &\; 
t_0\cdot f_t(\Box)t_0 -2\nu\, \Big( m\,\big(2\tilde{\rho}_0+2\tilde{\pi}_0l_0-v\,l_0l_0\big) + \\ 
&\qquad\quad
+\tilde{\rho}_0\tilde{f}_\rho(\Box)\tilde{\rho}_0 +\big(\tilde{\pi}_0-vl_0\big)\tilde{f}_\pi(\Box)\big(\tilde{\pi}_0-vl_0\big)\Big)\Big] \ . \nonumber
\end{align}
From this renormalized action we can retrieve Ward identities that are completely equivalent to those in~\eqref{WIv}, with inverted roles for $v$ and $m$ (and $\nu$ going into $-\nu$).

To conclude the discussion, we would like to provide an explicit expression for the two-point correlator of $\ImOt$, where the massless Goldstone pole should be found. For $v=0$, that in alternative quantization corresponds to purely spontaneous breaking, and $\nu=1/2$, corresponding to the dimension of the boundary operator equal to $1/2$, the equation of motion~\eqref{eqL} becomes
\begin{equation}\label{solvablenu}
M''(z) - \left(k^2+ {m^2}{z}^{-1}\right) M(z) =0 \ ,
\end{equation}
where $M=z\partial_zL$. This equation can be analitically solved, and, if we impose boundary conditions such that the solution is not exploding in the deep bulk, we obtain the following well-behaved function
\begin{equation}
M(z)= C\; z\, e^{-\sqrt{k^2\,}z}\: \sfU\bigg[1+\frac{m^2}{2\sqrt{k^2\,}},\,2;\,2\sqrt{k^2\,}\,z\bigg] \ ,
\end{equation}
where $\sfU[a,b;x]$ is the Tricomi's hypergeometric function.

From the constraint~\eqref{eqAz} we get
\begin{equation}
\tilde{\pi}_0= -\frac{1}{m} \left.k^2 M \right|_{z^0} \ ,
\end{equation}
where $\left.M\right|_{z^0}$ is the constant term in the small~$z$ expansion. Similarly, from the equation of motion~\eqref{eqL} we can express the gauge invariant combination involving $\pi_0$ in the following way:
\begin{equation}
\pi_0-ml_0 = -\frac{1}{m} \left. M' \right|_{z^0} \ .
\end{equation}
Then we can derive the final expression for the correlator
\begin{align}
\big\langle\ImOt(k)\,\ImOt(-k) \big\rangle & = i\,\tilde{f}(k^2) = i\,\frac{\pi_0-ml_0}{\tilde{\pi}} \\
& =
	-\frac{i}{k^2} \left[ \sqrt{k^2\,}-m^2\left( 2\gamma_{EM} +\ln\!\big(2\sqrt{k^2\,}\big) + \psi_{(2)}\bigg[1+\frac{m^2}{2\sqrt{k^2\,}}\bigg]\right)\right]\ , \nonumber
\end{align}
where $\gamma_{EM}$ is the Euler-Mascheroni constant, and $\psi_{(2)}[x]$ is the di-gamma function. Using the expansion $\psi_{(2)}[1+x]\simeq \ln (x) + 1/(2x) + {\mathcal O}(1/x^2)$ for large $x$, one verifies that both the linear term $\sim |k|$ and the logarithmic term $\sim \ln|k|$ in the numerator of the equation above cancel in the $k\to 0$ limit. In this way, the low energy behavior of this correlator exhibits the expected Goldstone massless pole, namely
\begin{equation}
\big\langle\ImOt\,\ImOt \big\rangle \approx i\, \frac{2m^2}{k^2}\: \big(\gamma_{EM} +\ln m\big) \ .
\end{equation}
{We have thus confirmed the presence of the Goldstone boson, in addition to the constraints imposed by the Ward identities.}

\section{Discussion}

In the present work we have verified from the holographic point of view that in the strict large $N$ limit spontaneous symmetry breaking can occur in two dimensions~\cite{Witten:1978qu}. Indeed, considering the minimal AdS$_3$/CFT$_2$ setup in which symmetry breaking can be produced, we have retrieved the canonical Ward identities as they appear in higher dimensions~\cite{Argurio:2015wgr}. Nevertheless, the way to get this result involves subtleties and peculiarities which are specific to two dimensions, and can be regarded as a premonition of the fact that spontaneous breaking is ruled out as soon as one moves away from strict infinite $N$.  The most crucial subtlety is that we have to renormalize the gauge field in the alternative quantization, if we want it to properly source a conserved current and then recover the correct Ward identities for the breaking of a global symmetry on the boundary.

{We can consider quantum corrections to the result that we have obtained, taking inspiration from~\cite{Anninos:2010sq}, and compute a bulk tadpole correction to the scalar profile. It would presumably reproduce the infrared divergence which is responsible for preventing the vacuum expectation value, mirroring a similar field theory computation. Such a quantum effect, by the holographic correspondence, is equivalent to a $1/N$ correction in the boundary theory.}

On the other hand, we can think of the question directly in field theory, considering the canonical example of a complex scalar field with a mexican-hat potential, as in \cite{Ma:1974tp}. In $1\!+\!1$ dimensions, the large quantum fluctuations of the phase would prevent the selection of a specific ground state around the circle at the minimum of the potential. However, if we add an arbitrarily small (but finite) explicit breaking, this would select a particular ground state, and act as a regulator for the infra-red divergence, making such vacuum stable under quantum fluctuations. Hence, for explicit breaking parametrically smaller than the spontaneous one, we expect (even at finite $N$) a mode that is hierarchically lighter than the rest of the spectrum, and whose mass is linear in the explicit breaking parameter, in accordance with the renowned Gell-Mann--Oakes--Renner relation~\cite{GellMann:1968rz}. So, if there are no Goldstone bosons in two dimensions, there definitely {should be} pseudo-Goldstone bosons in two dimensions, {and we have just provided a holographic description of the latter.}

\vskip 3em

{\noindent{\bf Acknowledgments.} {We would like to thank Matteo Bertolini, Massimo Bianchi, Alejandra Castro, St\'ephane Detournay, Jos\'e Edelstein, Jelle Hartong, Nabil Iqbal, Daniele Musso, and Massimo Porrati for useful discussions and exchange of opinions.}
This research has been supported in part by IISN-Belgium (convention 4.4503.15). R.A. is a Senior Research Associate of the Fonds de la Recherche Scientifique-F.N.R.S. (Belgium). The work of G.G. has been supported by CONICET through grant PIP 0595-13 and a NSF-CONICET bilateral cooperation grant. G.G. and J.A.S.-G. thank the hospitality of Universit\'e Libre de Bruxelles and Solvay Institutes of Belgium, where substantial part of this project has been done. J.A.S.-G. acknowledges support from a Spanish FPI fellowship (FPA2011-22594), from MINECO (FPA2014-52218-P), Xunta de Galicia (GRC2013-024) and FEDER.
}

\end{document}